\documentclass[12pt]{article}
\usepackage[utf8x]{inputenc}
\usepackage{amssymb}
\usepackage{amsmath}
\usepackage{graphicx}
\usepackage{hyperref}

\newcommand{\eq}{\begin{equation}}
\newcommand{\eqx}{\end{equation}}
\newcommand{\eqs}{\begin{equation*}}
\newcommand{\eqsx}{\end{equation*}}
\newcommand{\eqn}{\begin{eqnarray}}
\newcommand{\eqnx}{\end{eqnarray}}
\newcommand{\alg}{\begin{align}}
\newcommand{\algx}{\end{align}}

\newcommand{\f}[2]{\frac{#1}{#2}}

\newcommand{\cor}[1]{\left\langle{#1}\right\rangle}

\newcommand{\lm}{\lambda}
\newcommand{\Lm}{\Lambda}

\newcommand{\sg}{\sigma}

\newcommand{\Dl}{\Delta}

\newcommand{\om}{\omega}
\newcommand{\gm}{\gamma}
\newcommand{\kap}{\kappa}
\newcommand{\kaps}{\kappa_{spin}}
\newcommand{\eps}{\varepsilon}
\newcommand{\qqqq}{\quad\quad\quad\quad}

\newcommand{\RR}{{\mathbb{R}}}

\DeclareMathOperator{\arccosh}{arccosh}

\newcommand{\tr}{\mbox{\rm tr}\,}

\newcommand{\nn}{{\cal N}}
\newcommand{\WW}{{\cal W}}

\newcommand{\tx}{\tilde{x}}
\newcommand{\tz}{\tilde{z}}

\title{On correlation functions of operators dual to classical spinning string states}

\author{Romuald A. Janik\thanks{e-mail: {\tt romuald@th.if.uj.edu.pl}},\ 
Piotr Surówka\thanks{e-mail: {\tt surowka@th.if.uj.edu.pl}},\  
Andrzej Wereszczyński\thanks{e-mail: {\tt wereszcz@th.if.uj.edu.pl}}}

\date{Institute of Physics\\
Jagiellonian University\\
ul. Reymonta 4\\ 
30-059 Krak\'ow\\
Poland}

\begin{document}

\maketitle

\begin{abstract}
We explore how to compute, classically at strong coupling, correlation functions 
of local operators corresponding to classical spinning string states.
The picture we obtain is of `fattened' Witten diagrams, the evaluation of
which turns out to be surprisingly subtle and requires a modification of 
the naive classical action due to a
necessary projection onto appropriate wave functions.
We examine string solutions which compute the simplest case of 
a two-point function and reproduce the right scaling with the anomalous
dimensions corresponding to the energies of the associated spinning string
solutions. We also describe, under 
some simplifying assumptions, how the 
spacetime dependence of a conformal three-point correlation function
arises in this setup. 
\end{abstract}

\vfill
\pagebreak

\section{Introduction}

In a conformal field theory the full dynamical information is contained in 
two ingredients: the spectrum of conformal weights and the coefficients of 
three point functions of primary operators (OPE coefficients). All higher 
point correlation functions are expressible in terms of these 
data\footnote{This has been exploited in detail for two-dimensional 
CFT's \cite{BPZZ}, but should also hold for higher dimensional 
theories \cite{Fradkin}.}.

In the case of $\nn=4$ SYM theory, thanks to the developments of integrability 
\cite{BPR}-\cite{BES}
in the AdS/CFT correspondence \cite{adscft,GKP,Witten}, we have a very 
detailed quantitative understanding 
of the structure of the spectrum of conformal weights (i.e. anomalous dimensions 
of operators), being very close to an explicit solution \cite{TBAAF,TBAGKV,TBABomb}. 
However, there is almost no information on the OPE coefficients apart from 
early works on 3-point functions of protected operators dual to supergravity 
modes \cite{cubic1,cubic2}, short operators like Konishi \cite{KonishiOPE}, 
various operators in the BMN limit as most recently considered in \cite{Russo}.

This imbalance comes from the fact that in order to compute generic OPE coefficients 
one has to consider three point functions of primary operators which are dual 
to \emph{massive} string states, for the calculation of which we do not have 
appropriate technical tools. Indeed since the calculational methods for
computing correlation functions in the AdS/CFT correspondence are based on
Green's functions and an action for the corresponding fields \cite{GKP,Witten},
they have been so far restricted to operators dual to supergravity fields.
For operators dual to massive string modes, this route appears to be closed
due to the absence of a workable closed string field theory in $AdS_5 \times S^5$.

The main motivation for the present paper is to address this issue. 
Specifically, we would like to develop a setup where one may use classical 
(or even perhaps semi-classical) methods for computing correlation functions 
of operators dual to (massive) classical string states at strong coupling.
In this paper we will concentrate on two-point functions, which should 
reproduce the known anomalous dimensions equal to the energies of the
corresponding spinning string solutions in $AdS_5 \times S^5$ \footnote{For a stringy
classical analysis of the calculation of 
correlation functions in the BMN limit see \cite{Yoneya}-\cite{Yoneya2}.}. In addition, 
we will show how the standard conformal spacetime dependence of three point 
functions arises from the same setup. We will also describe various aspects 
of the calculations which were quite surprising for us.

The plan of this paper is as follows: first we will describe some motivation 
for the existence of a classical regime for a class of correlation functions 
involving massive string states, then we will proceed to review a point 
particle example first in flat space and then in $AdS_5$. In section 4 we
will describe various difficulties that one faces when trying to extend
these calculations to the case of classical string states. In section 5
we will formulate our prescription and apply it, in the following section,
to compute two point functions of operators dual to spinning string solutions.
We will describe generic solutions in $S^5$ and a specific example with spins
also in the $AdS_5$ factor. Finally, in section 7, we will describe how 
the spacetime dependence of a 3-point correlation function arises from 
this setup. We close the paper with a summary and an outlook. 

\section{Motivation}

The conformal weights (anomalous dimensions) of gauge theory operators are 
extracted from the string side of the AdS/CFT correspondence 
by computing the energies of string states in $AdS_5 \times S^5$,
and using the identification between the symmetries on both sides of the 
correspondence. 
The energy of a string state in $AdS_5 \times S^5$ measured w.r.t. global AdS 
time is an eigenvalue of the operator
\eq
\f{1}{2} (P_0+K_0)
\eqx
This operator is conjugate (by a  \emph{nonunitary} similarity transformation) 
to  $i$ times the dilatation operator. Therefore the spectrum of conformal 
weights of $\nn=4$ SYM coincides with the spectrum of energies of a superstring 
in $AdS_5 \times S^5$.

In particular, at strong coupling, there is a class of operators with large 
R-charges or spins which are dual to spinning string classical solutions in 
$AdS_5 \times S^5$ \cite{GKP2,FT,TseytlinReview}. Consequently, their anomalous 
dimensions behave like
\eq
\Delta \sim  \sqrt{\lm} \cdot F(charges)
\eqx
On the CFT side, we know that the two-point correlation function of the 
corresponding operator has to have the form
\eq
\label{e.corrfunc}
\cor{O(0) O(x)} = \f{const.}{|x|^{2\Delta}}
\eqx
For consistency, it should therefore be possible to also reconstruct the same 
anomalous dimension from a direct 
computation of the correlation function (\ref{e.corrfunc}) on the string side 
of the AdS/CFT correspondence. Moreover, we expect that a \emph{classical} 
computation should suffice, as the relevant correlation function, for operators 
dual to spinning string states, scales as
\eq
\cor{O(0) O(x)} = \f{const.}{|x|^{2\Delta}} \sim e^{-2\sqrt{\lm} \cdot F(charges) \cdot 
\log |x|}
\eqx
at strong coupling, which is a typical form of an Euclidean action of 
a classical string solution. Moreover, the coefficient of $\log |x|$ should 
exactly coincide with the energy of the quite different Minkowskian spinning 
string located near the center of global $AdS_5 \times S^5$. 
For technical reasons we will consider these correlation
functions in Minkowski signature (with spacelike separation of the operator 
insertion points) which will be easily continued to Euclidean spacetime.

We expect, therefore, to have a mapping from classical spinning string solutions 
rotating around the center of Minkowskian global $AdS_5 \times S^5$ 
to certain solutions in the Poincare patch which realize 
the two point functions for any choice of the positions of the operator 
insertions on the boundary. This new classical solution should approach 
the boundary at exactly these points.

\begin{figure}

\hfill\includegraphics[height=4cm]{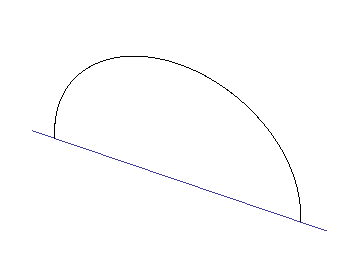} \hfill
\includegraphics[height=4cm]{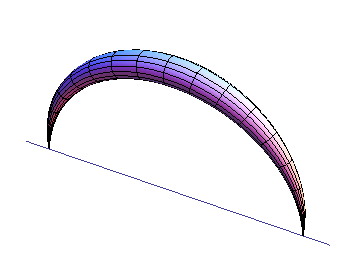}\hfill{}

\caption{Standard Witten diagram for a two point function representing an 
exchange of a supergravity field and a `fattened' version representing 
the exchange of a classical spinning string state.}
\end{figure}

For higher point functions we may expect to have a structure of `fattened' 
Witten diagrams \cite{Witten} (shown in fig.~1 for the case of two point 
functions), where the type of solution approaching the relevant point on 
the boundary represents a kind of `classical' vertex operator for the spinning 
string state\footnote{It would be interesting to compare with \cite{vertex},
building upon \cite{TseytlinVertex}.}. 
In section 7 we will describe the spacetime structure 
of a three-point function in this setup.

\section{A point particle example}

A prototype for our considerations is the derivation of a point particle Green's 
function from the worldline formalism. One can obtain the scalar field Green's 
function $G(x,y)$ by evaluating a path integral over the worldlines of 
a particle between the two points $x,y$ weighted with the Polyakov (particle) 
action and performing an integral over the modular parameter of the trajectories. 
This calculation has been carried out in detail in \cite{Polchinski}, 
here we will just discuss its classical limit.
Let us start with Euclidean signature.

Then one has to perform the path integral
\eq
\int [de] [dx^\mu] \exp \left( -\f{1}{2} \int_0^1 (e^{-1} \dot{x}^2+e m^2) dt \right)
\eqx
with the boundary conditions $x^\mu(0)=x^\mu$, $x^\mu(1)=y^\mu$.
In the above expression, $e$ is the einbein field related to the metric on 
the worldline. One can 
use diffeomorphisms to set $e$ to a constant $s$ which becomes the modular 
parameter of the worldline trajectory. Equivalently, we may keep the worldline
metric equal to unity and extend the range of the worldline time coordinate.
In this way one gets
\eq
\int_0^\infty ds \int [dx^\mu] (measure) \exp \left( -\f{1}{2} \int_0^s 
\left(\dot{x}^2+ m^2\right) dt \right)
\eqx 
Let us now evaluate the above expression by saddle point. The solution with appropriate boundary
conditions is $x^\mu(t)=(y^\mu-x^\mu)t/s+x^\mu$, giving for the Polyakov action
\eq
S_P=\f{1}{2} \left( \f{|x-y|^2}{s}+m^2 s \right)
\eqx
Performing a saddle point w.r.t. the modular parameter gives the correct large 
distance asymptotics of the flat space scalar Green's function. 
\eq
G(x,y) \sim e^{-m |x-y|}
\eqx
A careful, exact evaluation of the path integral reproduces the exact scalar 
Green's function \cite{Polchinski}. 
Repeating the same computation in Minkowski signature, we would start from 
the action
\eq
\exp \left( \f{i}{2} \int_0^s \left(\dot{x}^2- m^2\right) dt \right) 
\eqx
and then, depending on whether the separation $\Dl x^\mu \equiv y^\mu-x^\mu$ 
is spacelike or timelike we will arrive at a real or complex saddle point 
($s=|\Dl x|/m$ or $s=-i|\Dl x|/m$) giving the
following standard asymptotics of the Feynman propagator in flat space
\eq
G(x,y) \sim e^{-m |x-y|}  \qqqq \text{or} \qqqq G(x,y) \sim e^{-i m |x-y|}
\eqx
It is clear that the spacelike separation case essentially coincides with the 
Euclidean result.

Let us now consider the point particle in $AdS_5$, and specialize to the case
relevant for a two point function. This calculation corresponds e.g. to
a heavy KK supergravity mode which is dual to a protected operator of large
(classical) dimension.

For the calculation it is convenient to use the Poincare coordinates of 
Euclidean $AdS_5$:
\eq
ds^2_{AdS_5^E} = \f{dx^2+dz^2}{z^2}
\eqx
The two point function in the AdS/CFT correspondence is essentially 
the Green's function, regularized by moving the operator insertion points
by $\eps$ into the bulk i.e. to $(0,\eps)$ and $(x,\eps)$ with $\eps \to 0$.

The relevant action for a point particle of mass $m^2_{AdS}$ is
\eq
S_P= \f{1}{2} \int_{-\f{s}{2}}^{\f{s}{2}} d\tau 
\biggl\{ \f{\dot{x}^2 +\dot{z}^2}{z^2} + m^2_{AdS} \biggr\} 
\eqx
where $s$ is the modular parameter of the worldline. 
The equations of motion in AdS are
\eqn
\partial_\tau \left( \f{2\dot{x}}{z^2} \right) &=& 0 \\
\partial_\tau \left( \f{2\dot{z}}{z^2} \right) &=& \f{-2}{z^3} (\dot{x}^2 +
\dot{z}^2)
\eqnx
These equations are solved by
\eqn
\label{e.adssols}
x(\tau) &=& R\, \tanh \kap \tau +x_0  \nonumber \\
z(\tau) &=& R\, \f{1}{\cosh \kap \tau}
\eqnx
which is, in fact, a specific parametrization of a geodesic $(x(\tau)-x_0)^2+z^2=R^2$
in $AdS_5$.
With this solution the AdS part of the action simplifies to
\eq
\f{\dot{x}^2 +\dot{z}^2}{z^2}=\kap^2
\eqx
We now have to impose the boundary conditions
\eq
\biggl(x(-s/2),z(-s/2)\biggr)=(0,\eps) \qqqq \biggl(x(s/2),z(s/2)\biggr)=(x,\eps) 
\eqx
Since $\eps$ is very small, $\kap s$ has to be large and we get
\eq
x=2R \tanh \frac{\kappa s}{2}  \;\; \Rightarrow \;\; R\sim \f{x}{2}  \;\;\;\; x_0 \sim R
\eqx
and
\eq
\kap=\f{2}{s} \arccosh \f{R}{\eps} \sim \f{2}{s}\log \f{x}{\eps}
\eqx
\eq
S_P=\f{1}{2} \left( \kap^2 +m^2_{AdS}  \right) s=
\f{1}{2} \left( \f{4}{s^2}\log^2 \f{x}{\eps} +m^2_{AdS}  \right) s
\eqx
We now have to perform integration over the modular parameter $s$, which we do 
by saddle point. The saddle point solution is
\eq
s=\f{2\log \f{x}{\eps}}{m_{AdS}}
\eqx
leading to
\eq
e^{-S_P} = e^{- 2 m_{AdS} \log \f{x}{\eps}} =
\left( \f{|x|}{\eps} \right)^{-2m_{AdS}} 
\eqx
In this way we recovered the standard relation between particle 
masses in $AdS$ and operator dimensions in the large 
mass limit $\Delta=m_{AdS}+corrections$.\\

If we pass to Minkowski signature in the AdS case there are some new features
in comparision to the flat space case. For spacelike separation along the $x$ 
coordinate, the action becomes
\eq
S_P= \f{1}{2} \int_{-\f{s}{2}}^{\f{s}{2}} d\tau 
\biggl\{ \f{\dot{x}^2 +\dot{z}^2}{z^2} - m^2_{AdS} \biggr\} 
\eqx
Here the solution of the equations of motion is as in the Euclidean case. The 
only difference is the sign of $m^2_{AdS}$ which gives rise to an imaginary
saddle point in $s$, which, when substituted back to the action compensates
the prefactor $i$ in $\exp(i S_P)$ and gives the purely real 
answer $|x|^{-2m_{AdS}}$.

For timelike separation, say along the $t$ direction, the action is
\eq
S_P= \f{1}{2} \int_{-\f{s}{2}}^{\f{s}{2}} d\tau 
\biggl\{ \f{-\dot{t}^2 +\dot{z}^2}{z^2} - m^2_{AdS} \biggr\} 
\eqx
Now, in contrast to the flat space case, it turns out that there is no 
appropriate real trajectory approaching the
boundary and one is forced to consider an analytically continued complex
trajectory which again gives rise to the real exponent $|t|^{-2m_{AdS}}$.

The obvious way to obtain the Green's function for massive string states is 
to consider a cylinder amplitude for the closed string, similar to the one 
considered in \cite{Polchinski,Marshakov}, and evaluate it by saddle point, 
ensuring that the propagating string state corresponds to a classical state 
with appropriate angular momenta. 
However it turns out that this procedure is quite subtle as we will discuss in
the following section.

\section{Problems with extension to strings}

As a simple example which illustrates the problems encountered in
extending the point particle calculation of the previous section to the
closed string context, let us consider the case of the flat space
cylinder amplitude described in \cite{Polchinski,Marshakov}. In particular,
for the case of pointlike boundary conditions for the endpoints of the cylinder,
the flat space amplitude (for the bosonic closed string) evaluated
in Euclidean signature takes the form:
\eq
\label{e.flatcyl}
\int_0^\infty \f{ds}{s^{13}} e^{4\pi s} \prod (1-e^{-4\pi n s})^{-24} 
e^{-S_P(\Dl x)}
\eqx
where $s$ is the modular parameter of the cylinder, $S_P(\Dl x)$ is the classical 
action for a pointlike straight string stretched 
between point $x^\mu$ and point $x^\mu+\Dl x^\mu$ i.e.
\eq
S_P(\Dl x)=-\f{\Dl x^2}{4\pi s}
\eqx
Once we expand the product in the preexponential factor, we will 
get~\cite{Polchinski}
\eq
\int_0^\infty \f{ds}{s^{13}} \sum_{N=0}^\infty d_N e^{-4\pi s m^2_N} 
e^{-\f{(\Dl x)^2}{4\pi s}}=
\sum_{N=0}^\infty d_N \int \f{d^{26}p}{(2\pi)^{26}} \f{e^{ip\Dl x}}{p^2+4m_N^2}
\eqx
in which we at once recognize a summation over all string states propagating
along the cylinder. Each summand is in fact just the worldline representation
of the Green's function of the corresponding string state, exactly as dicussed 
in the previous section. So the interpretation of each term in the cylinder
amplitude is perfectly natural and in line with expectations.

Suppose now that we were to try to extract from this setup the Green's functions of
a highly excited massive string state which would be almost classical --- 
a rotating string with large angular momenta, a typical example of which would be
the following Minkowskian solution \cite{FT}:
\eq
\label{e.rotflat}
x^1+ix^2=a_1 \sin n_1 \sg e^{in_1 \tau} \qqqq
x^3+ix^4=a_2 \sin n_2 (\sg+\sg_0) e^{in_2 \tau}
\eqx
supplemented by
\eq
x^0=\kap \tau \quad\quad x^L=\f{\Dl x}{s}\tau
\eqx
It is convenient to interpret this state treating the coordinates where 
the rotation takes space as being in some `compactified' space, while the
coordinates $x^0,x^L$ would be the `physical' ones. In this way the similarity to
the setup of spinning strings in $\RR \times S^5 \subset AdS_5 \times S^5$ is
greatest.

Once we try to extract the Green's function corresponding to this particular
solution from the cylinder amplitude, we encounter a series of difficulties.
Firstly, the calculation (\ref{e.flatcyl}) was performed in Euclidean signature, 
so the above
solution would not be a solution of the Euclidean equations of motion. At 
the least it would have to be complexified, thus losing in this case any 
similarity with 
a rotating string. So suppose that to overcome this obstacle we consider the
cylinder amplitude in Minkowski signature. 
Now at least the state (\ref{e.rotflat}) is a solution to the equations of
motion. But then, we would expect it, being a solution to the classical 
equation of motion, to contribute through the
exponent of the classical \emph{action} in the cylinder formula. This would
lead to an incorrect result as the Minkowskian action is quite different from 
the Minkowskian \emph{energy} which is effectively responsible for the $mass^2$ 
of the string state.

So let us go back to the original Euclidean formula (\ref{e.flatcyl}) and examine
more closely how the correct Green's functions were reproduced. There, 
the mass terms did not arise from the classical action but rather from the
fluctuation determinant. So they were associated to \emph{eigenfunctions} of
the Laplace operator and these, almost by chance, turned out to be related to 
solutions of the original Minkowskian equations of motion.

If we were to try to apply a similar Euclidean calculation to the case of 
$AdS_5 \times S^5$, and try to extract contributions related to 
classical spinning strings we would fail, as we would not even have the picture
of determinants and eigenfunctions since the Polyakov action in $AdS_5 \times S^5$
is nonlinear. Moreover we would not expect spinning string states to be associated
to any kind of `small fluctuations'.

In the next section we will go back to Minkowskian signature and show
how one does overcome the obstacles presented here. Let us emphasize that what
we are after is a purely path integral derivation without any recourse to the
operator formulation of the cylinder amplitude as $\tr e^{s (L_0+\bar{L}_0)}$.
The reason for going this long-winded way is that we would like to obtain a
(semi-)classical formulation applicable also to higher point functions and not
only to the cylinder.

\section{Semiclassical propagator revisited}

The basic problem with obtaining the correct answer for the propagator of
a certain classical string mode is that apparently, the path integral
formula for a cylinder amplitude in the classical limit  
is necessarily dominated by
\eq
e^{i S_{class}}
\eqx 
while the effective mass is governed not by the action
but by the (classical) energy.

The resolution of this puzzle is in fact very simple, although we did not
find it spelled out anywhere in the path integral literature.

Consider first a quantum mechanical system with one degree of freedom. Then, 
let us concentrate on a certain semi-classical state, eigenfunction of the
hamiltonian, and consider its quantum
time evolution. Its wave function must of course evolve according to its energy, 
which should be roughly, for the state in question, equal to its classical value.
However, the semiclassical propagator for the system is governed only by the
\emph{actions} of classical trajectories. This is in essence the exact counterpart
of our string cylinder puzzle.

But now, the answer is obvious. We have to convolve the semiclassical propagator
with the wavefunction of the state that we are interested in. Since we are
considering an almost classical state, we may use the leading WKB form of the
wave function:
\eq
\int dx_i \underbrace{e^{i\int^{x_i} p(x) dx}}_{\Psi(x_i)} \cdot 
e^{i S_{class}[x_i,x_f,T]}
\eqx
where $S_{class}[x_i,x_f,T]$ is the action evaluated for some classical 
trajectory. In the classical limit
we may evaluate the convolution with the wavefunction by saddle point obtaining
\eq
p(x_i)+ \f{\partial S_{class}[x_i,x_f,T]}{\partial x_i}=p(x_i)-p=0
\eqx
where $p$ is the initial momentum of the classical trajectory dominating 
the propagator.
Now since consequently both $x_i$ and $p$ have to coincide with $x_i$ and $p(x_i)$ of our 
classical state in question, the trajectory determining the propagator has to 
be the same as the one entering the wavefunction. The result will be
\eq
\exp \left\{i\int^{x_i} p(x) dx\right\} \cdot\exp\left\{i S_{class}[x_i,x_f,T]
\right\}
\eqx 
which can be rewritten as
\eq
\exp \left\{i S_{class}[x_i,x_f,T]\right\} \cdot \exp 
\left\{-i \int_{x_i}^{x_f} p(x) dx\right\} \cdot \exp 
\left\{i \int^{x_f} p(x) dx\right\}
\eqx
The first two factors combine just to the standard energy phase
\eq
S_{class}[x_i,x_f,T]-\int_{x_i}^{x_f} p(x) dx = \int_0^T \left( L-p \dot{x}
\right) dt =-E_{class} T
\eqx
where we used the fact that the classical trajectory entering the propagator 
and the wavefunction have to coincide.
In this way we recover the right result $\Psi(x_f) e^{-i E_{class} T}$ purely from 
classical (saddle point) considerations.

In the case of many degrees of freedom, one has to use the Hamilton-Jacobi equation
for stationary states in order to generate the WKB wavefunction and the same 
reasoning goes through. The same also holds
in the field theory limit.

Now its obvious that we should repeat the same procedure for the string cylinder 
amplitude.
We should project the amplitude onto the contribution of the classical
state that we are interested in by convolving with the relevant wave function.
The convolution ensures that the classical solution for the cylinder should
coincide with the classical state in question. Moreover, moving the wavefunction
`across' the cylinder will transform the contribution of that trajectory from
$\exp (i S_{class}[in,out,s])$ to $\exp(-i E_{class} s)$. 

There is still one more subtlety which arises for the string cylinder amplitude.
Up till now we looked at the flat space example as a product of ``physical space'',
where the arguments of the Green's function are located, and ``compactified space''
where the string was rotating (analog of the $S^5$ in $AdS_5 \times S^5$). However
in order to correctly extract Green's functions of string states with nontrivial
rotation also in the  ``physical space'', we have to subtract from the wave function
the zero mode (which will enter the arguments of the Green's function). Concretely,
we will have
\eq
\label{e.subtgen}
\exp (i S_{class}[in,out,s]) \cdot \exp\left( -i\int d\sg d\tau\, (\pi-\pi_0)
\cdot (\dot{x}-\dot{x}_0) \right)
\eqx
where $\pi_0$ and $\dot{x}_0$ are the zero mode parts of the canonical momentum
and velocity i.e.
\eq
\pi_0(\tau)\equiv\f{1}{2\pi} \int_0^{2\pi} d\sg\, \pi(\tau,\sg)
\qqqq
\dot{x}_0(\tau)\equiv\f{1}{2\pi} \int_0^{2\pi} d\sg\, \dot{x}(\tau,\sg)
\eqx
This notion is ambiguous in $AdS_5$ spacetime and we will make a concrete proposal
in this case in the following section, where we will apply the above procedure
to evaluate two-point correlation functions of operators dual to spinning 
string states.

\section{Two point correlation functions -- spinning strings}

Let us now analyze two point correlation functions of operators which are dual to
classical spinning strings. We will first consider a string rotating 
in\footnote{After the completion of this paper we became aware that a very similar 
construction for strings spinning in $S^5$ was performed in \cite{tsuji}.} $S^5$ 
and then give an example of a spinning string rotating both in $AdS_5$ and 
in $S^5$. Let us emphasize, that of course
the outcome of the calculation is obvious both from the field theory point of 
view, and from our knowledge of the worldsheet string hamiltonian.
Our objective is, however, to perform the calculation directly using purely 
classical methods so that the generalization of the setup to higher point 
correlation functions would be clear.

\subsection*{A simple example -- circular rotating string in $\RR \times S^5$}

Let us parametrize the metric on $S^5$ as
\eq
ds^2_{S^5} = d\gm^2+\cos^2 \gm d\phi_3^2+\sin^2\gm \left(d\psi^2 + \cos^2\psi 
d\phi_1^2+ \sin^2 \psi \phi_2^2 \right)
\eqx
The simplest spinning string is the circular rotating string with two equal
spins. It is contained in the subspace
\eq
\gm= \f{\pi}{2} \qqqq \phi_3=0
\eqx
with the nontrivial fields being
\eq
\psi=\sg \qqqq \phi_1=\phi_2=\om\tau   \qqqq \text{and} \quad 
\tau_{AdS}= \kaps \tau 
\eqx
Here $\tau_{AdS}$ is the global AdS time, while $\om$ is related to the
R-charges through $J_1=J_2\equiv J\equiv \sqrt{\lm}j =\sqrt{\lm}\om/2$. The spinning string 
solution obeys Virasoro constraints which in this case reduce to 
$\kaps^2=1+\om^2$. The global AdS energy of the solution is thus
\eq
E_{AdS}=\kaps=\sqrt{1+\om^2}=\sqrt{1+4j^2}
\eqx

Let us now compute the two point function of the corresponding gauge theory 
operator. As in the case of the point particle we will stay in Minkowski signature
and consider the insertion points to be spacelike separated.
We have to evaluate a cylinder amplitude with the Polyakov action
\eq
\label{e.adscyl}
S_P=-\f{\sqrt{\lm}}{4\pi} \int_{-\f{s}{2}}^{\f{s}{2}} d\tau \int_0^{2\pi} d\sg
\biggl\{ -\f{\dot{x}^2 +\dot{z}^2}{z^2} +\text{$S^5$ part} \biggr\}
\eqx
The classical solution would be a solution of the equations of motion for any 
value of the modular parameter $s$ with the boundary conditions 
\eq
\label{e.bccor}
\biggl(x(-s/2),z(-s/2)\biggr)=(0,\eps) \qqqq \biggl(x(s/2),z(s/2)\biggr)=(x,\eps) 
\eqx
where we used the fact that there does not need to be any $\sg$ dependence for
the AdS part of the solution. Therefore $x(\tau)$ and $z(\tau)$ are exactly given 
by the solutions (\ref{e.adssols}). Since we are dealing with a cylinder amplitude,
we do not impose Virasoro constraints. The $S^5$ part of the solution is just
the $S^5$ part of the circular rotating string
\eq
\psi=\sg \qqqq \phi_1=\phi_2=\om\tau 
\eqx
Evaluating the Polyakov action for this configuration gives
\eq
\exp \left\{ i \f{\sqrt{\lm}}{2} \left(\kap^2+ \underbrace{(\om^2-1)}_{\text{$S^5$ action}}
\right) s \right\} 
\eqx
However, as explained before we have to include the effect of the wavefunction of
the rotating string state, which will change the $S^5$ action into its energy 
obtaining
\eq
\exp \left\{ i \f{\sqrt{\lm}}{2} \left(\kap^2- \underbrace{(\om^2+1)}_{\text{$S^5$ energy}}
\right) s \right\} 
\eqx
Taking into account the formula for $\kap$ we get
\eq
\label{e.circular}
\exp \left\{ i \f{\sqrt{\lm}}{2} \left( \f{4}{s^2}\log^2 \f{x}{\eps} -(\om^2+1) \right) s 
\right\} 
\eqx
The remaining saddle point in $s$ is just as for the case of the point 
particle described in section~3 and we obtain the correct two-point correlation
function
\eq
\cor{O(0)O(x)} =\f{1}{|x|^{2\sqrt{\lm} \sqrt{1+4j^2}}}
\eqx
One can check that the cylinder string solution at the (complex) saddle point 
in $s$ satisfies Virasoro constraints (similarly as in \cite{Zarembo}).

\subsection*{Generic spinning strings on $\RR \times S^5$}

The above analysis of the circular rotating string can be readily extended
to generic spinning string states in $\RR \times S^5$. In this case it is
convenient to write the $S^5$ part of (\ref{e.adscyl}) as
\eq
-\dot{X}^2+{X'}^2 + \Lm (X^2-1)
\eqx
As for the circular string, the AdS energy (which corresponds to the anomalous
dimension) of the spinning string solution is
\eq
E_{AdS}=\kaps
\eqx
where $\kaps$ of the spinning string is related by the Virasoro
constraint to the \emph{worldsheet energy} of the $S^5$ part of the solution,
namely
\eq
\kaps^2=\dot{X}^2+{X'}^2
\eqx
But this is exactly what we need to obtain the correct answer for the two-point 
correlation function. After using the wavefunction to project on the desired 
state, the analog of (\ref{e.circular}) becomes
\eq
\exp \left\{i \f{\sqrt{\lm}}{2} \left( \f{4}{s^2}\log^2 \f{x}{\eps} -E_{AdS}^2 \right) s 
\right\} 
\eqx
which leads to the correlation function
\eq
\cor{O(0)O(x)} =\f{1}{|x|^{2\sqrt{\lm} E_{AdS}}}
\eqx

\subsection*{Spinning string on $AdS_5 \times S^5$}

Let us now consider a more complicated example of a spinning string with
angular momenta both in the $S^5$ and $AdS_5$ factors. Here we will have to 
deal with
the complication that the nontrivial spinning motion of the string may interfere
with the `bending of the solution' necessary for the string to approach the
given operator insertion points on the boundary. Also, for the same reason, 
finding the solution directly will not be as trivial as in the preceeding cases.

We will not consider here the most general solution but rather a simple example which
exhibits all of the discussed features -- a circular spinning string with $J=S$ 
\cite{TR}.

The original spinning string solution has the following form in global 
coordinates on $AdS_5$:
\eqn
\label{e.twospin}
Y_0 &=& \cosh \rho_0 \sin \kaps \tau \nonumber\\
Y_1 &=& \sinh \rho_0 \cos(\om \tau+\sg) \nonumber\\
Y_2 &=& \sinh \rho_0 \sin(\om \tau+\sg) \nonumber\\
Y_3=Y_4 &=& 0 \nonumber\\
Y_5 &=& \cosh \rho_0 \cos \kaps \tau 
\eqnx
with the solution on $S^5$ being
\eqn
X_1 &=& \cos(w \tau-\sg) \nonumber\\
X_2 &=& \sin(w \tau-\sg) \nonumber\\
X_i &=& 0 \qqqq\qqqq \text{for $i>2$}
\eqnx
Equations of motion imply $\om=\sqrt{1+\kaps^2}$. The charges\footnote{Here we 
always extract factors of $\sqrt{\lm}$.} of the solution are $J=w$, 
$S=\om \sinh^2 \rho_0$ and
\eq
\label{e.enspinsj}
E=\kaps \cosh^2 \rho_0=\kaps +\f{\kaps S}{\sqrt{1+\kaps^2}}
\eqx
The off-diagonal Virasoro constraint requires $S=J$ while the remaining diagonal
one gives
\eq
\label{e.diagvir}
1-\kaps^2+J^2+\f{2S}{\sqrt{1+\kaps^2}}=0
\eqx
which, together with (\ref{e.enspinsj}), expresses the energy in terms of 
the spin $S=J$.

We will now use this solution to construct an appropriate solution in the Poincare
patch.  We will not impose the Virasoro constraint (\ref{e.diagvir}) and 
keep $\kaps$ arbitrary. Then we will perform the (nonunitary) transformation which
exchanges $H=\f{1}{2}(P_0+K_0)$ with $iD$, $Y_0 \to i Y_4$ and $Y_4 \to i Y_0$,
and finally we will set $\kaps \to i \kap$. In this way we obtain a solution to
the equations of motion which differs from (\ref{e.twospin}) by $Y_0=0$, 
$Y_4=\cosh \rho_0 \sinh \kap \tau$ and $Y_5=\cosh \rho_0 \cosh \kap \tau$. This
solution has a very transparent meaning in Poincare coordinates:
\eqn
\label{e.twospinvert}
t &=& 0 \nonumber\\
x_1 &=& \tanh \rho_0 \cos(\om \tau+\sg) e^{\kap \tau} \nonumber\\
x_2 &=& \tanh \rho_0 \sin(\om \tau+\sg) e^{\kap \tau} \nonumber\\
x_3 &=& 0 \nonumber\\
z &=& \f{1}{\cosh \rho_0} e^{\kap \tau} 
\eqnx
It represents a rotating string ``emitted'' from the origin at the boundary at
$\tau=-\infty$ and propagating into the bulk with $\om=\sqrt{1-\kap^2}$.

As a side remark, let us note that
if we were to insert $\kap =-i\kaps$ we would get a \emph{complex} solution which
has the correct value of \emph{purely imaginary} eigenvalue of the dilatation
operator. However, the nice spacetime picture would be lost. Going to Euclidean
signature on the worldsheet would cure the behaviour in the $z$ direction but
would mess up the spinning string structure. We will therefore always perform
computations in Minkowski space and only take a (complex) saddle point in the
final answer.

Now, in order to find the cylinder solution which approaches two given points on
the boundary it is enough to perform a special conformal transformation
\eq
x^\mu \to \frac{x^\mu + b^\mu (x^2+z^2)}{1+2 xb +b^2 (x^2+z^2)} \qqqq
z \to \frac{z}{1+2 xb +b^2 (x^2+z^2)}
\eqx
on the solution (\ref{e.twospinvert}). Let us choose $b^\mu=(0,1/R,0,0)$, so
that the separation of the operator insertion points interferes with the 
rotation of the string. The resulting solution looks now more complicated with
the nontrivial $AdS_5$ coordinate fields given by
\eqn
x_1 &=& \f{\tanh \rho_0 \cos(\om \tau+\sg) e^{\kap \tau}+
\f{1}{R} e^{2\kap \tau}}{1+\f{2}{R} \tanh \rho_0 \cos(\om \tau+\sg)
e^{\kap \tau}+\f{1}{R^2} e^{2\kap \tau}} \nonumber\\
x_2 &=& \f{\tanh \rho_0 \sin(\om \tau+\sg) e^{\kap \tau}}{
1+\f{2}{R} \tanh \rho_0 \cos(\om \tau+\sg)
e^{\kap \tau}+\f{1}{R^2} e^{2\kap \tau}} \nonumber\\
z &=& \f{\f{1}{\cosh \rho_0} e^{\kap \tau} }{
1+\f{2}{R} \tanh \rho_0 \cos(\om \tau+\sg)
e^{\kap \tau}+\f{1}{R^2} e^{2\kap \tau}}
\eqnx
We now have to impose the boundary conditions (\ref{e.bccor}) for the string 
cylinder. In contrast to the $S^5$ examples, there will be some $\sg$ 
dependence, but since we are interested
in the limit $\eps \to 0$ it can be neglected as we will need to have 
$e^{\kap s/2}$ either very large or very small. Asymptotically we obtain
\eq
\label{e.kapsj}
\kap=\f{2}{s} \log \f{R}{\eps \cosh \rho_0} \longrightarrow 
\f{1}{s} \log \f{R^2}{\eps^2}
\eqx
where we absorbed\footnote{This seems like a very natural choice of 
normalization, but we do not have an intrinsic justification for it. It is nevertheless 
necessary for obtaining the correct answer.} 
the $\cosh \rho_0$ into $\eps$.

The classical action of this cylinder solution is
\eq
iS_{class}=i\f{\sqrt{\lm}}{2} (\kap^2 + w^2-1) s
\eqx
We now have to perform the subtraction (\ref{e.subtgen}). Here we face 
the problem that for
generic curved spacetimes there is no unique way to define the zero modes, as 
the answer would be different in different coordinate systems. In the case of 
$AdS_5$, we may however single out a prescription by requiring that it respects 
the $SO(2,4)$ symmetry of the background. So we propose to use the global $Y^A$
coordinates to define the zero modes of the string namely
\eq
\dot{Y}^A_0 \equiv \f{1}{2\pi} \int_0^{2\pi} d\sg \, \dot{Y}^A(\sg,\tau)
\eqx
Due to the fact that $SO(2,4)$ acts linearly on the $Y^A$'s, the zero mode of
the transformed solution will be the transformation of the zero mode.

Let us now evaluate the correction term coming from convolution with 
the wavefunctions. From (\ref{e.subtgen}) we get
\eq
i\cdot \text{correction}=-i \f{\sqrt{\lm}}{2} (2\sinh^2 \rho_0-2\kap^2 
\sinh^2 \rho_0+2w^2)s
\eqx
which gives
\eq
\exp \left\{ i\f{\sqrt{\lm}}{2} (\kap^2+2\sinh^2 \rho_0 (\kap^2-1) -(1+w^2))s
\right\}
\eqx
Finally, let us extremize w.r.t. $s$. To this end we have to express
the above expression in terms of physical charges of the string state. Moreover
we will use (\ref{e.kapsj}) to evaluate the saddle point w.r.t. $\kap$ instead of
$s$. In this way we get for the exponent
\eq
i\f{\sqrt{\lm}}{2} \left( \kap-2S \f{\sqrt{1-\kap^2}}{\kap}-\f{1+J^2}{\kap} 
\right) \log \f{R^2}{\eps^2}
\eqx
The saddle point w.r.t. $\kap$ yields
\eq
\label{e.sadpoint}
1+\kap^2+J^2+\f{2S}{\sqrt{1-\kap^2}}=0
\eqx
which, for $\kap=i\kaps$, exactly coincides with the Virasoro condition of the
original spinning string solution (\ref{e.diagvir}). The cylinder amplitude 
at the saddle point $\kap=i\kaps$ thus gives
\eq
\exp \left\{ -\sqrt{\lm} \left( \kaps+\f{\kaps S}{\sqrt{1+\kaps^2}} \right) 
\log \f{R^2}{\eps^2} \right\} = \left( \f{\eps^2}{R^2} \right)^{\sqrt{\lm} 
\left( \kaps+\f{\kaps S}{\sqrt{1+\kaps^2}} \right)}
\eqx
where we used the saddle point equation (\ref{e.sadpoint}) to rewrite
the answer in a form identical with the formula for the energy of the
corresponding spinning string state (\ref{e.enspinsj}).

\subsection*{The setup for higher point correlation functions}

It is now roughly clear how in principle one should be able to compute a higher 
point correlation function. One should find a Minkowskian classical solution which
approaches the insertion points of the operators in question with the 
same asymptotic behaviour as for the well-understood two point correlation 
functions. The legs of the string solution should be joined together somewhere
in the bulk. The convolution with the wave functions should change the lagrangian
into hamiltonian densitities (up to the extraction of the zero modes), 
but the transported wavefunctions will now
be put in the vicinity of the junction point/cycle. The understanding of
the structure of this string joining in $S^5$, which is after all the main 
motivation for the present work, is a very interesting but complex problem which we
leave nevertheless for future research. Finally one has to extremize w.r.t. 
the modular parameters and the position of the juncture.

In this paper we will just demonstrate, under some
simplifying assumptions, how the standard spacetime dependence of a three point
function in a conformal field theory arises from the above procedure.

\section{Three point functions}

\begin{figure}

\hfill\includegraphics[height=4cm]{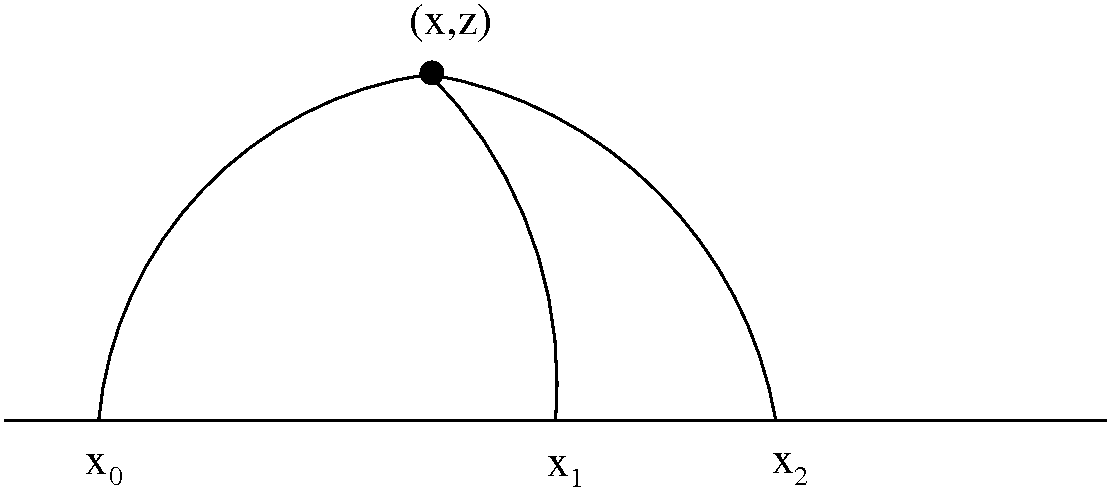} \hfill{}

\caption{A Witten diagram for a three point correlation function.}
\end{figure}

Let us now consider three point correlation functions of operators dual to spinning 
string states rotating in $S^5$.
We expect a string configuration corresponding to the Witten diagram shown in 
fig.~2.

We will perform the calculation assuming that the solutions are undeformed away 
from the string junction. This should at least be true for the case of pointlike 
strings. In this paper we will not study the interaction vertex in more detail 
leaving this for future work. In particular we assume that the three external 
states are such that the three cylinders \emph{may} be joined
together at the string junction.

Let us consider the Witten diagram configuration shown in fig.~2. Each of the 
legs will be a cylinder solution with modular parameter $s_i$ extending from 
the point $(x_i,\eps)$ near the boundary to the string junction point $(x,z)$. 
The action of the whole system will have
to be extremized w.r.t. the modular parameters $s_i$, as well as the coordinates 
of the string junction.

Using the formulas of previous sections we see that the modular parameters of 
the cylinders with the above boundary conditions can be expressed through 
the parameters of the solutions as
\eq
s_i=\f{1}{\kap_i} \left( \arccosh \f{R_i}{z}-\arccosh \f{R_i}{\eps} \right)
\eqx
Since the $R_i$'s are independent of $\kap_i$,
instead of extremizing w.r.t. $s_i$ we may extremize w.r.t. $\kap_i$.
The exponent appearing in the amplitude for the three cylinders, after taking 
into account convolution with the wave functions is
\eq
\exp{\WW}\equiv \exp\left\{i \f{\sqrt{\lm}}{2} \sum_i (\kap_i^2-E_i^2) 
\f{1}{\kap_i} \left( \arccosh \f{R_i}{z}- \arccosh \f{R_i}{\eps} \right)\right\}
\eqx
Taking the saddle point w.r.t. the $\kap_i$'s gives
\eq
\WW=\sum_i \Delta_i \left( \arccosh \f{R_i}{z}-\arccosh \f{R_i}{\eps} \right)
\eqx
where we put $\Dl_i=\sqrt{\lm} E_i$.
We now have to evaluate the parameter $R_i$ in terms of the boundary conditions. 
A quick calculation gives
\eq
R_i^2 = \f{(z^2+(x-x_i)^2)^2}{4 (x-x_i)^2} 
\eqx
The exponent $\WW$ can be further simplified to
\eq
\label{e.wewant}
\WW[x_0,x_1,x_2]=\sum_i \Dl_i \log \f{z\eps}{z^2+(x-x_i)^2}
\eqx
where we have explicitly indicated the dependence on the insertion points.
This is just a linear combination of the $AdS$ invariant distances between 
the points $(x_i,\eps)$ and $(x,z)$ weighted with the conformal dimensions of 
the external states\footnote{This is not surprising, as from the $AdS_5$ point 
of view the computation becomes essentially a point-particle one.}. Now we have 
to find the saddle point w.r.t. $x$ and $z$. This is 
very difficult, if not impossible, to do explicitly. However we may use 
the conformal symmetry of $AdS_5$ to find the exact dependence of the answer 
on the positions of the operator insertion points $x_0,x_1,x_2$. For simplicity 
we will just consider these points to lie along one coordinate axis so that 
the endpoints of the string cylinders are at 
$(x_0,\eps)$, $(x_1,\eps)$, $(x_2,\eps)$, with $x_0,x_1,x_2$ being scalars.

Consider first the specific choice $(\tx_0=0,\eps)$, $(\tx_1=1,\eps)$, $(\tx_2=-1,\eps)$.
The exponent $\WW$ will be extremized by some specific values of $z=\tz$ and $x=\tx$:
\eq
\WW[\tx_0,\tx_1,\tx_2]=\sum_i \Dl_i \log \f{\tz\eps}{\tz^2+(\tx-\tx_i)^2}
\eqx
Now let us take a conformal transformation which transforms the $\tx_i$ into 
$x_i$. This induces an isometry of $AdS_5$ which transforms
\eq
(\tx_i,\eps) \longrightarrow (x_i,\eps_i)
\eqx
Since the Polyakov action is invariant we have
\eq
\WW[\tx_0,\tx_1,\tx_2]=\sum_i \Dl_i \log \f{z\eps_i}{z^2+(x-x_i)^2}
\eqx
where the transformed $z$ and $x$ automatically solve the saddle point equations for
$\WW[x_0,x_1,x_2]$.
But this is almost exactly what we want for the general case. Indeed, comparision 
with (\ref{e.wewant}) gives
\eq
\label{e.fin}
\WW[x_0,x_1,x_2]=\WW[0,1,-1]-\sum_i \Dl_i \log \f{\eps_i}{\eps}
\eqx
So the whole spacetime dependence sits in the last term which is easy to
evaluate.
In order to complete the calculation, we need the explicit form of the $AdS_5$ 
isometry.
For the case at hand, it may be constructed from a composition of a dilatation, 
a special
conformal transformation and a translation. Explicitly we have
\eqn
x &\longrightarrow & \f{\lm x+b \lm^2 (x^2+z^2)}{1+2\lm b x+b^2 \lm^2 (x^2+z^2)}+x_0 \\
z  &\longrightarrow & \f{\lm z}{1+2\lm b x+b^2 \lm^2 (x^2+z^2)}
\eqnx
The parameters $\lm$ and $b$ are given in our case as
\eq
\lm = -\f{2x_{01}x_{02}}{x_{12}}   \qqqq b=\f{x_{10}+x_{20}}{2x_{10}x_{20}}
\eqx
Plugging the above into (\ref{e.fin}), we obtain 
\eq
\f{1}{|x_{10}|^{\Dl_0+\Dl_1-\Dl_2} |x_{20}|^{\Dl_0+\Dl_2-\Dl_1} 
|x_{12}|^{\Dl_1+\Dl_2-\Dl_0} }
\eqx
which is the expected spacetime dependence of the three point correlation 
function in a conformal field theory.

\section{Summary and Outlook}

In this paper we have addressed the problem of computing correlation functions
of operators dual to classical spinning string solutions. This can be done
directly on the string side of the AdS/CFT correspondence using a classical
computation. We have found that special care has to be taken when projecting
on the string state in question. Convolution with semiclasssical wave functions
modifies the classical action of the string solution by certain correction
terms which are crucial in order to obtain the correct answer. The same methods
should also apply to other versions of the AdS/CFT correspondence like 
\cite{ABJM}. It would also be interesting to investigate to what extent these
methods could be extended to less symmetric examples of AdS/CFT.

We analyzed in detail the case of two point functions which are evaluated
through a cylinder amplitude and a saddle point extremization w.r.t. the modular
parameter. Again, the classical action has to be modified here. The two point
correlation function computations, for the examples considered in the present
paper, reproduce the correct scaling governed by the anomalous dimension
equal to the energy of the associated spinning string state. Moreover, 
the string solution entering the two point correlation function computation,
provides the asymptotic behaviour of classical string configurations which 
would enter higher point correlation functions involving the operator 
in question.  

For the case of three point correlation functions we have shown how the
expected spacetime dependence arises from our setup under some simplifying
assumptions. We leave further investigation of the three point functions
for future work.

The results obtained in the present paper lead to numerous directions for
further research.

Even staying within the setting of two point correlation functions, it would
be interesting to give a precise proof of the equivalence for all classical
finite gap solutions. Also it would be very nice to develop 
the formulation beyond the strict classical limit and include quadratic
fluctuations. This might be especially interesting for short operators, like
Konishi, where the formulation involving fluctuations around a geodesic
close to the boundary seems to be quite distinct from a short string living
in the center of $AdS_5$ in an almost flat geometry. This point of view might 
thus lead to
a cross check of the currently available string computations at strong coupling
\cite{AFKonishi,RTKonishi} and the fit to numerical Y-system 
results \cite{GKVKonishi}.

The most interesting further directions to explore would be connected with
the understanding of the classical solutions associated with three point
correlation functions. These would have the topology of a sphere with three
holes (effectively punctures in the $\eps \to 0$ limit). The solutions in 
the vicinity of these punctures should approach the operator insertion points 
on the boundary with the asymptotic behaviour characteristic of the given 
operator, which can be read off from the known two point correlation 
function solution.
The properties and conditions for the existence of such solutions remain, for
the moment, a completely open problem. However, the machinery of integrability
should certainly be applicable here.

\bigskip

\noindent{\bf Acknowledgments:} This work was supported by Polish science funds 
as research projects N N202 105136 (2009-2011) (RJ and PS), N N202 239637 (PS) and 
N N202 126735 (2008-2010) (AW). RJ was supported by Marie Curie ToK 
KraGeoMP (SPB 189/6.PRUE/2007/7). RJ thanks IPMU, Tokyo and IPhT Saclay
for hospitality and support during work on this project. PS thanks
Max-Planck-Institut f\"ur Physik in Munich for hospitality and support
during the completion of this work. RJ would like to thank Sergey Frolov and Ryo
Suzuki for interesting discussions. Finally, we thank Arkady Tseytlin for comments and informing us of Ref. \cite{tsuji}.

\end{document}